\documentclass[english,twocolumn, prl]{revtex4}
\usepackage[latin9]{inputenc}
\usepackage{graphicx}
\usepackage{amssymb}
\usepackage{esint}

\usepackage{babel}

\begin{document}

\title{Universal dephasing in a chiral 1D interacting fermion system}

\author{Clemens Neuenhahn}

\author{Florian Marquardt}

\address{Department of Physics, Arnold Sommerfeld Center for Theoretical Physics,
and Center for NanoScience, Ludwig Maximilians Universit\"at M\"unchen,
Theresienstr. 37, 80333 Munich, Germany}
\begin{abstract}
We consider dephasing by interactions in a one-dimensional chiral
fermion system (e.g. a Quantum Hall edge state). For finite-range
interactions, we calculate the spatial decay of the Green's function
at fixed energy, which sets the contrast in a Mach-Zehnder interferometer.
Using a physically transparent semiclassical ansatz, we find a power-law
decay of the coherence at high energies and zero temperature ($T=0$),
with a universal asymptotic exponent of 1, independent of the interaction
strength. We obtain the dephasing rate at $T>0$ and the fluctuation
spectrum acting on an electron.
\end{abstract}
\maketitle
Studying the loss of quantum coherence is important both for fundamental
reasons (quantum-classical transition, measurement process, equilibration)
and with regard to possible applications of quantum mechanics (interferometry,
quantum information processing). 

Dephasing of electrons in Luttinger liquids is interesting as an example
of a non-perturbative, strongly correlated model case \cite{1983_ApelRice_LLdephasing,2002_LeHur_LLdephasing,2005_LeHur_ElectronFractionalization,2005_Mirlin_DephasingLLweakLoc,2006_LeHur_LifetimeLL,2008_04_GutmanGefenMirlin_NonEquilibriumLL}.\textbf{
}In contrast, the situation for (spinless) chiral interacting fermion
systems, such as edge states in the integer quantum Hall effect (QHE),
seems to be clear. Within the standard ansatz of pointlike interactions,
an interacting chiral model is only a Fermi gas with a renormalized
velocity. Recently though it was realized that such models may present
interesting physics if finite-range interactions are considered \cite{2007_ChalkerGefen_MZ}
(cf. also\textbf{ }\cite{1993_MedenSchoenhammer_SpectrumLL}). This
research is motivated by recent studies of dephasing in QHE Mach-Zehnder
interferometers, both by intrinsic interactions \cite{2001_SeeligBuettiker_MZDephasing,2006_04_LawFeldmanGefen_FQHE_MZ,2007_05_NederMarquardt_NJP_NonGaussian,2006_09_Sukhorukov_MZ_CoupledEdges,2007_ChalkerGefen_MZ,2007_11_NederGinossar_ShotNoiseDephasing,2008_01_Shenja_TwoChannelDephasing}
and external baths \cite{2004_Marquardt_MZ_PRL,2004_10_Marquardt_MZQB_PRL,2005_Foerster_MZ_FCS,2006_04_MZQB_Long,2006_07_MZ_DephasingNonGaussianNoise_NederMarquardt,2007_05_NederMarquardt_NJP_NonGaussian}.
Remarkable experiments \cite{2003_Heiblum_MachZehnder,2006_01_Neder_VisibilityOscillations,2006_07_MZ_DephasingNonGaussianNoise_NederMarquardt,2008_02_Strunk_MZ_Coherence_FillingFactor,2008_03_Roche_CoherenceLengthMZ}
have revealed novel effects at high bias voltages, which is the regime
we are going to study. 

At low energies and temperatures, chiral interacting fermions form
a Fermi liquid and are fully coherent at $T=0$ and $\epsilon=\epsilon_{F}$.
It was found that the features at intermediate energies depend on
the details of the interaction potential \cite{1993_MedenSchoenhammer_SpectrumLL,PhysRevB.60.4571,2007_ChalkerGefen_MZ}.
However, here we study the coherence of interacting chiral fermions
at \emph{high} energies (higher than the cutoff for the interaction
potential). Our central result is that (at $T=0$) there is a \emph{universal}
power-law decay of coherence with propagation distance, where the
exponent is \emph{independent} of interaction strength. This is in
contrast to physical expectation, where decoherence should grow with
increasing coupling. We identify the reason behind this as a subtle
cancellation between increasing interaction strength and decreasing
density fluctuations in the sea of other electrons. We will derive
this first within a semiclassical ansatz that is later shown to be
exact at high energies, comparing it to bosonization. We will discuss
deviations from the leading behaviour and the situation at $T>0$.
The result is particularly remarkable since usually universal behaviour
is confined to the low-energy regime.

\emph{The model}. \textendash{} We consider fermions in one dimension,
propagating chirally at speed $v_{F}$ and interacting via a potential
$U(x-x')$: %
\begin{figure}
\includegraphics[width=1\columnwidth]{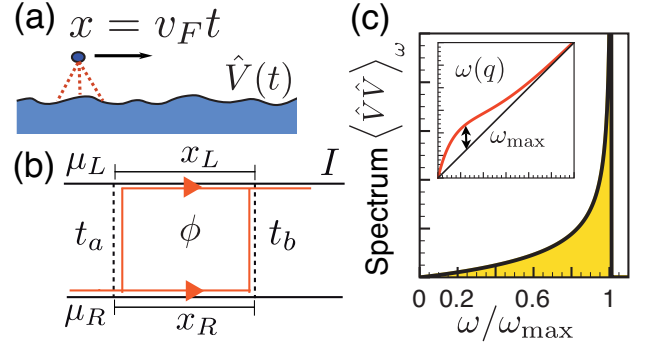}

\caption{\label{specfig}(a) A single electron propagating at high energies
feels a fluctuating potential $\hat{V}(t)$, as it interacts with
the sea of other electrons. (b) Scheme of the Mach-Zehnder interferometer
setup. (c) The fluctuation spectrum (at $T=0$ and $\alpha>0$). Inset:
plasmonic dispersion relation. \label{specfig}}

\end{figure}

\begin{eqnarray}
\hat{H} & = & \sum_{k}v_{F}k\hat{c}_{k}^{\dagger}\hat{c}_{k}+\nonumber \\
 &  & \frac{1}{2}\int dx\, dx'\,\hat{\psi}^{\dagger}(x)\hat{\psi}^{\dagger}(x')U(x-x')\hat{\psi}(x')\hat{\psi}(x)\,,\end{eqnarray}
where $\hat{\psi}(x)=L^{-1/2}\sum_{k}\hat{c}_{k}e^{ikx}$ are the
fermion operators, the normalization volume $L$ tends to infinity
in the end, $k\in2\pi L^{-1}\mathbb{Z}$, and $k\geq k_{c}$, with
a cutoff $k_{c}$ that drops out of the results. We have set $\hbar=1$.
After bosonization, the Hamiltonian is diagonal:

\begin{equation}
\hat{H}=\sum_{q>0}\omega(q)\hat{b}_{q}^{\dagger}\hat{b}_{q}+\mu\hat{N}.\label{eq:Hbos}\end{equation}
The bosonic operators $\hat{b}_{q}$ of Eq.~(\ref{eq:Hbos}) describe
the density fluctuations $\hat{\rho}(x)\equiv\hat{\psi}^{\dagger}(x)\hat{\psi}(x)-\bar{\rho}$,
where $\bar{\rho}$ is the mean density:

\begin{equation}
\hat{\rho}(x)=\sum_{q>0}\sqrt{\frac{q}{2\pi L}}(\hat{b}_{q}e^{iqx}+{\rm h.c.}).\end{equation}
The plasmonic dispersion relation depends on the interaction potential's
Fourier transform, $U_{q}=\int dx\, e^{-iqx}U(x)$:

\begin{equation}
\omega(q)=v_{F}q\left[1+\frac{U_{q}}{2\pi v_{F}}\right]\,.\end{equation}
Here $v_{F}$ is the velocity at $q\rightarrow\infty$ and we define
$\tilde{v}\equiv v_{F}+U_{q\rightarrow0}/(2\pi)=v_{F}(1+\alpha)$.
The dimensionless interaction strength is $\alpha\equiv U_{0}/(2\pi v_{F})$.
$U_{q}$ is assumed to decay beyond some scale $q_{c}$.

\emph{Interferometry}. \textendash{} To probe the electrons' coherence,
we imagine an electronic Mach-Zehnder interferometer {[}Fig.$\,$\ref{specfig}(b){]}
\cite{2001_SeeligBuettiker_MZDephasing,2003_Heiblum_MachZehnder},
i.e. two chiral wires connected by small tunnel couplings $t_{a}$
and $t_{b}$ at two {}``beam splitters'' (quantum point contacts).
This permits to express the current to leading order in the tunnel
coupling \cite{2006_09_Sukhorukov_MZ_CoupledEdges,2007_ChalkerGefen_MZ},
via the single-particle Green's functions (GF) in the wires. This
is possible under the assumption that there are no interactions between
the wires (and therefore no vertex corrections in the result), which
is reasonable due to their spatial separation. The quantity of interest
is the visibility $\mathcal{V}$, i.e. the contrast of the current
interference pattern that is displayed when changing the magnetic
flux $\phi$. We define $\mathcal{V}\equiv(I_{{\rm max}}-I_{{\rm min}})/(I_{{\rm max}}+I_{{\rm min}})$,
where $I_{{\rm max}}={\rm max}_{\phi}I(\phi)$. In contrast to \cite{2007_ChalkerGefen_MZ},
we write $\mathcal{V}$ in terms of the GF in energy-space: $G^{>}(x,\epsilon)$
is the Fourier transform of $G^{>}(x,t)=-i\langle\hat{\psi}(x,t)\hat{\psi}^{\dagger}(0,0)\rangle$.
It gives the amplitude for an electron injected at energy $\epsilon$
to propagate coherently a distance $x$. This yields (at $T=0$): 

\begin{equation}
\mathcal{V}=\frac{2|t_{a}t_{b}^{\ast}|}{|t_{a}|^{2}+|t_{b}|^{2}}\cdot\frac{\left|\int_{0}^{\delta\mu}d\epsilon\; G_{L}^{>}(x_{L},\epsilon)G_{R}^{<}(-x_{R},\epsilon-\delta\mu)\right|}{(2\pi)^{2}\int_{0}^{\delta\mu}d\epsilon\;\nu_{L}(\epsilon)\nu_{R}(\epsilon-\delta\mu)}\,.\end{equation}
There are contributions from all electrons inside the voltage interval,
$\epsilon=0\ldots\delta\mu$, where $\delta\mu=\mu_{L}-\mu_{R}=q_{e}V>0$
is the bias between the left (L) and the right (R) interferometer
arm. $G_{L,R}$ are the bulk GF's {[}at $\mu=0${]} for particles
($>$) and holes ($<$), where $G^{<}(x,\epsilon)=G^{>*}(x,-\epsilon)$.
At $T=0$ one obtains the tunneling density of states from $2\pi\nu(\epsilon)=\left|G^{>}(x=0,\epsilon)\right|+\left|G^{<}(0,\epsilon)\right|$.
For $x_{L}=x_{R}=x$, the decay of visibility is thus determined by
the GF decay to be discussed in the following. 

\emph{Decoherence of a high-energy electron}. \textendash{} We employ
a physically intuitive semiclassical ansatz for the GF's, that becomes
exact in the limit of high energies, as we will confirm later by comparing
it to bosonization. Electrons at high energies $\epsilon\gg v_{F}q_{c}$
propagate at the speed $v_{F}$. Scattering by a few multiples of
$q_{c}$ will not bring them near the Fermi energy, so Pauli blocking
is unimportant. The visibility at high bias voltage is dominated by
these electrons. The sea of other electrons produces a fluctuating
potential $\hat{V}(t)$ acting on such a high-energy electron at its
classical position $x=v_{F}t$. It is obtained by convoluting the
density with the interaction potential {[}Fig.~\ref{specfig}(a){]}:\begin{equation}
\hat{V}(t)=\int dx'\, U(x'-v_{F}t)\hat{\rho}(x',t).\label{eq:VUrho}\end{equation}
As known from bosonization, the fluctuations of $\hat{\rho}$ are
purely Gaussian. The ansatz assumes the electron to pick up a random
phase from the potential fluctuations $\hat{V}(t)$. As a result,
its non-interacting GF $G_{0}^{>}$ is multiplied by the average of
the corresponding phase factor: $G^{>}(x,\epsilon)=G_{0}^{>}(x,\epsilon)\cdot\exp(-F(x))$,
where\begin{eqnarray}
e^{-F(x)} & \equiv & \left\langle \hat{T}\exp\left[-i\int_{0}^{x/v_{F}}dt'\hat{V}(t')\right]\right\rangle \nonumber \\
 & = & \exp\left[-\frac{1}{2}\int_{0}^{x/v_{F}}dt_{1}dt_{2}\,\left\langle \hat{T}\hat{V}(t_{1})\hat{V}(t_{2})\right\rangle \right]\label{eq:e^-F}\end{eqnarray}
depends on the propagation distance $x$, but turns out to be energy-independent
in the high-energy limit discussed here. A related approach was introduced
both for electron dephasing in 1D ballistic wires by an external quantum
environment \cite{2004_10_Marquardt_MZQB_PRL,2006_04_MZQB_Long},
and for describing two coupled (non-chiral) Luttinger liquids \cite{2005_LeHur_ElectronFractionalization}
or 1D systems with a nonlinear dispersion relation \cite{2006_PustilnikGlazman_NonlinearDispersionRelationSpectrum,2008_ImambekovGlazman_EdgeSingularities}.
The form of $e^{-F(x)}$ is exactly the same as that for pure dephasing
of a qubit by quantum noise \cite{2004_10_Marquardt_MZQB_PRL,2006_04_MZQB_Long,2007_05_NederMarquardt_NJP_NonGaussian}.
The decay is determined by the fluctuation spectrum in the electron's
frame of reference, $\left\langle \hat{V}\hat{V}\right\rangle _{\omega}=\int dt\, e^{i\omega t}\left\langle \hat{V}(t)\hat{V}(0)\right\rangle $.
The magnitude of the GF (i.e. the electron's coherence) turns out
to decay as\begin{equation}
\frac{\left|G^{>}(x,\epsilon)\right|}{|G_{0}^{>}(x,\epsilon)|}=\exp\left[-\int_{-\infty}^{+\infty}\frac{d\omega}{2\pi}\frac{\sin^{2}(\omega x/2v_{F})}{\omega^{2}}\left\langle \left\{ \hat{V},\hat{V}\right\} \right\rangle _{\omega}\right],\label{eq:Ggreater}\end{equation}
where $\langle\{\hat{V},\hat{V}\}\rangle_{\omega}=\langle\hat{V}\hat{V}\rangle_{\omega}+\langle\hat{V}\hat{V}\rangle_{-\omega}$
denotes the symmetrized spectrum and $|G_{0}^{>}(x,\epsilon)|$ is
constant in the high-energy regime. From Eq.~(\ref{eq:VUrho}), we
obtain for the potential spectrum

\begin{equation}
\left\langle \hat{V}\hat{V}\right\rangle _{\omega}=\int\frac{dq}{2\pi}\,\left|U_{q}\right|^{2}\left\langle \hat{\rho}\hat{\rho}\right\rangle _{q,\omega+v_{F}q}\,,\label{eq:VVomega}\end{equation}
which derives from the Galileo-transformed spectrum of the density
fluctuations. We first focus on $T=0$, where $\left\langle \hat{\rho}\hat{\rho}\right\rangle _{q,\omega}=\theta(q)q\delta(\omega-\omega(q))$.
The spectrum has two distinct features (cf. Fig.~\ref{specfig}(c)).

At high frequencies, we obtain a singularity $\langle\{\hat{V},\hat{V}\}\rangle_{\omega}^{T=0}\propto1/\sqrt{\omega_{{\rm max}}-|\omega|}$
at the cutoff frequency $\omega_{{\rm max}}={\rm max}(\omega(q)-v_{F}q)$,
which is the maximum frequency in the Galileo-transformed plasmon
dispersion relation. This singularity arises since $\omega(q)\approx\omega(q^{*})+\omega''(q^{*})\cdot(q-q^{*})^{2}/2$
in the vicinity of $q^{*}$, where $\omega(q^{*})=v_{F}q^{*}+\omega_{{\rm max}}$. 

\begin{figure}
\includegraphics[width=1\columnwidth]{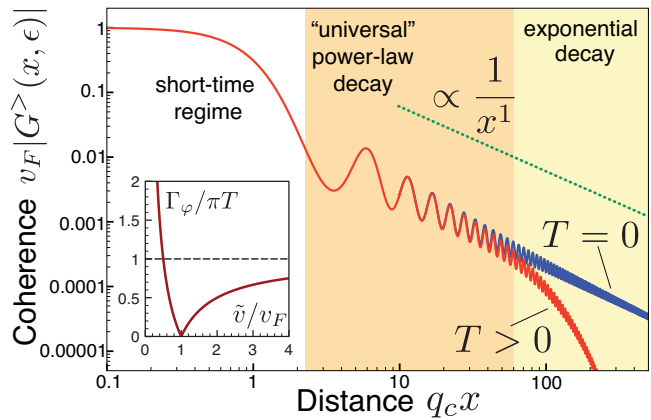}

\caption{\label{decayfig}The coherence of an electron propagating at high
energies in an interacting chiral system, as a function of propagation
distance. The non-interacting case would be $v_{F}\left|G^{>}(x,\epsilon)\right|\equiv1$.
The asymptotic exponent for the power-law decay is universally given
by $1$ (see dashed line). At $T>0$, one obtains an exponential decay
for large $x$ , with a decay rate $\Gamma_{\varphi}$ (inset). The
potential was $U_{q}=U_{0}e^{-|q/q_{c}|}$ with $U_{0}/v_{F}=2\pi\alpha=20$,
and $T/q_{c}v_{F}=0.01$.}

\end{figure}

At low frequencies $\omega\ll v_{F}q_{c}$, the spectrum increases
linearly in $\omega$, corresponding to {}``Ohmic'' noise, which
is ubiquitous in many contexts \cite{2000_Weiss_QuantumDissipativeSystems}.
Here, it derives from the interaction with 1D sound waves (plasmons).
For potentials that are smooth in real space (i.e. all the moments
of $\left|U_{q}\right|$ are finite), the leading low-$\omega$ behaviour
is determined by small $q$ in Eq.~(\ref{eq:VVomega}). The result
is :

\begin{equation}
\left\langle \left\{ \hat{V},\hat{V}\right\} \right\rangle _{\omega}^{T=0}=\frac{U_{q\rightarrow0}^{2}}{(\tilde{v}-v_{F})^{2}}\frac{|\omega|}{2\pi}\,=2\pi|\omega|.\label{VVom}\end{equation}

\[
\]
The prefactor of the spectrum turns out to be independent of the coupling
strength $\alpha$. This is in contrast to non-chiral Luttinger liquids,
where an Ohmic spectrum has been found with an interaction-dependent
prefactor \cite{2005_LeHur_ElectronFractionalization}. An increase
in interaction strength is canceled by stiffening the density fluctuations,
i.e. shifting them to higher frequencies in the co-moving frame, and
thereby decreasing their magnitude. This translates into a universal
power-law decay for the GF at large $x$ :

\begin{equation}
|G^{>}(x,\epsilon)|\propto\frac{1}{x^{1}}.\label{eq:decay}\end{equation}
More precisely, we claim that asymptotically the exponent becomes
$1$: $\lim_{x\rightarrow\infty}-\ln|G^{>}(x,\epsilon)|/\ln x=1$.
While here the cancellation of $\alpha$ is unexpected, a similar
effect is known for Nyquist noise, where the electron charge cancels
at low $\omega$ due to screening. Note the contrast to dephasing
by an external bath, where the decay gets weaker for lower coupling,
and also to the coupling-dependent exponents in a Luttinger liquid.
This central result is illustrated in Fig.$\,$\ref{decayfig}, based
on Eq.~(\ref{eq:Ggreater}). The power-law decay reflects the Anderson
orthogonality catastrophe, where the many-body state of the 'other'
electrons evolves depending on the path of the given electron. The
oscillations are due to the cutoff in $\langle\hat{V}\hat{V}\rangle_{\omega}$.
Its amplitude depends on $\alpha$ (see below), but it vanishes for
large $x$. These oscillations can be understood as 'coherence revivals',
where the entanglement with the environment is partly undone at certain
times, in the manner of 'quantum eraser' experiments. %

In order to understand how the non-interacting limit is recovered
($\alpha=0$, where $|G^{>}(x,\epsilon)|$ is constant as a function
of $x$), we have to discuss its range of validity. As the linear
slope in the spectrum applies only at $|\omega|\ll\omega_{{\rm max}}$,
we must require $\omega_{{\rm max}}x/v_{F}\gg1$. Since $\omega_{{\rm max}}$
vanishes with $\alpha$, the limiting regime is reached at ever larger
values of $x$ for $\alpha\rightarrow0$.

We now discuss the deviations from the leading low-$\omega$ behaviour
in $\langle\{\hat{V},\hat{V}\}\rangle_{\omega}$. These are due to
the contributions from large $q$ in (\ref{eq:VVomega}). E.$\,$g.
a potential $U_{q}=U_{0}e^{-(|q|/q_{c})^{s}}$ yields a sub-leading
contribution $\langle\{\hat{V},\hat{V}\}\rangle_{\omega}^{(sub)}=2\pi\left|\omega\right|/(s\ln(\left|\alpha\right|v_{F}q_{c}/|\omega|))$.
This turns into a term $s^{-1}\ln(\ln(|\alpha|q_{c}x))$ in $F(x)$,
yielding a slow logarithmic decay of the prefactor in Eq.~(\ref{eq:decay})
that can be understood as an asymptotically vanishing correction $s^{-1}\ln(\ln(|\alpha|q_{c}x))/\ln(x)\rightarrow0$
to the exponent $1$. The subleading oscillatory contribution to $F$
is $-C_{s}\sin(\omega_{{\rm max}}x/v_{F}+\pi/4)/\sqrt{2\pi\left|\alpha\right|q_{c}x}$,
with a numerical prefactor $C_{s}$ {[}e.g. $C_{1}=2\pi\sqrt{e}${]}.
In contrast, consider a potential that is non-smooth (i.e. $\int dq|U_{q}|q^{n}$
does not converge for some $n$). If $U_{q}=uq^{-n}$ for large $q$,
then we find an additional contribution $2\pi|\omega|(n-1)^{-1}$
{[}$n>1${]}. It modifies the leading behaviour of $\langle\{\hat{V},\hat{V}\}\rangle_{\omega}$
and changes the decay into $|G^{>}(x,\epsilon)|\propto1/x^{1+1/(n-1)}$.
The universal exponent is recovered as $n\rightarrow\infty$.

For $T>0$, the large-$x$ limit yields an exponential decay $|G^{>}(x,\epsilon)|\propto\exp[-\Gamma_{\varphi}x/v_{F}]$,
with 

\begin{equation}
\Gamma_{\varphi}=\pi T\left|1-\frac{v_{F}}{\tilde{v}}\right|=\pi T|1+\alpha^{-1}|^{-1}\,.\end{equation}
For $\alpha\rightarrow0$, this rate vanishes as $\Gamma_{\varphi}=\pi T|\alpha|$,
i.e. it is non-analytic in $U_{0}\propto\alpha$. Dephasing rates
linear in $T$ have also been found in non-chiral Luttinger liquids
\cite{2005_Mirlin_DephasingLLweakLoc,2005_LeHur_ElectronFractionalization,2006_LeHur_LifetimeLL}.
At large repulsion, $U_{0}\rightarrow+\infty$, we have the universal
result $\Gamma_{\varphi}\rightarrow\pi T$. For attractive interaction,
$\Gamma_{\varphi}$ diverges at the instability for $\alpha\rightarrow-1$,
where $\tilde{v}\rightarrow0$ gives rise to thermally excited low-frequency
modes.

Contrast this behaviour against pure dephasing of a qubit by Nyquist
noise. There, a power-law decay $t^{-\gamma}$ at $T=0$ implies a
decay rate $\Gamma_{\varphi}=\pi\gamma T$ for $T>0$. In the present
case, the Galileo transformation turns the lab-frame temperature $T$
into $T_{{\rm eff}}$ in the co-moving frame. We find $T_{{\rm eff}}=T\left|1-v_{F}/\tilde{v}\right|$
enters in the fluctuation-dissipation theorem relation $\langle\{\hat{V},\hat{V}\}\rangle_{\omega}^{T}=(2T_{{\rm eff}}/\left|\omega\right|)\langle\{\hat{V},\hat{V}\}\rangle_{\omega}^{T=0}$.
Only for large repulsion, we get $T_{{\rm eff}}\rightarrow T$, and
the universal power-law for $T=0$ turns into a universal decay rate
for $T>0$.

\begin{figure}
\includegraphics[width=1\columnwidth]{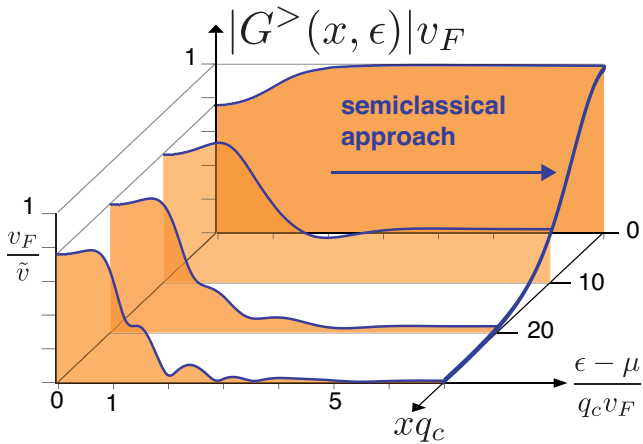}

\caption{\label{fig:GEX}The evolution of the Green's function with energy
$\epsilon$ of the injected electron at $T=0$, for various propagation
distances, according to bosonization {[}Eq.~(\ref{eq:GgreaterBOS}){]}.
The curve at the right corresponds to the semiclassical ansatz {[}Eq.~(\ref{eq:Ggreater}){]},
which is exact for high energies, as is evident in the figure. The
potential was $U_{q}=U_{0}e^{-(q/q_{c})^{2}}$ with $U_{0}/v_{F}=2\pi\alpha=2$.}

\end{figure}

\emph{Green's function from bosonization}. \textendash{} We employ
the standard connection \cite{1998_10_DelftSchoeller_BosonizationReview,2006_Giamarchi_Book}
between the bosonic phase field $\hat{\Phi}(x)=i\sum_{q>0}\sqrt{\frac{2\pi}{Lq}}e^{-aq}\left[\hat{b}_{q}(t)e^{iqx}-h.c.\right]$
and the fermion operators $\hat{\psi}(x)=\frac{\hat{F}}{\sqrt{2\pi a}}e^{ik_{F}x}e^{-i\hat{\Phi}(x)}$
(where $\hat{b}_{q}(t)=\hat{b}_{q}(0)e^{-i\omega_{q}t}$, $\hat{F}$
is the Klein factor, and $a\rightarrow0$ provides the regularization
at short distances). This yields the GF

\begin{eqnarray}
G^{>}(x,t) & = & \frac{-i}{2\pi a}e^{-i\mu(t-x/v_{F})}\cdot\nonumber \\
 &  & \exp\left[\left\langle \hat{\Phi}(x,t)\hat{\Phi}(0,0)\right\rangle -\left\langle \hat{\Phi}(0,0)^{2}\right\rangle \right]\,.\label{eq:GgreaterBOS}\end{eqnarray}
A numerical Fourier transform produces $G^{>}(x,\epsilon)$ (see Fig.$\,$\ref{fig:GEX}). The
dip near $\epsilon=0$ in the tunneling density $\propto|G^{>}(x=0,\epsilon)|$
is due to the renormalization of the velocity. The decay of the GF
with increasing $x$ is due to interaction-induced decoherence. Most
importantly, the decay at high energies (i.e. $\epsilon-\mu\gg v_{F}q_{c},\,\omega_{{\rm max}}$)
is reproduced exactly by the semiclassical approach (see Fig.~\ref{fig:GEX}).
This may be understood as follows: Evaluation of (\ref{eq:GgreaterBOS})
produces a broad, dispersing peak \cite{2007_ChalkerGefen_MZ} moving
with the renormalized velocity $\tilde{v}$. There is another, sharp
peak at $x=v_{F}t$. This is due to contributions from high frequencies
in the plasmon dispersion, and the evolution of its weight determines
the decay of $G^{>}(x,\epsilon)$ at high energies. That weight can
be obtained from bosonization(\ref{eq:GgreaterBOS}), evaluated at
$x=v_{F}t$, which turns out to be identical to the semiclassical
ansatz in Eq.~(\ref{eq:Ggreater}).\textbf{ }

In interferometry, these universal results determine the visibility
$\mathcal{V}$ for high bias voltage. At $T=0$ we obtain a decay
$\mathcal{V}\propto1/x^{2}$ independent of $\delta\mu$ at high bias
(note that $\mathcal{V}\rightarrow1$ for $\delta\mu\rightarrow0$,
as expected \cite{2007_ChalkerGefen_MZ}), and the exponential decay
for $T>0$ is transferred to $\mathcal{V}$ as well.

\emph{Conclusions}. \textendash{} The coherence of an electron moving
in a chiral system obeys a universal asymptotic power-law decay at
$T=0$, with an exponent $1$ independent of interaction strength,
for energies above the scale set by the interaction range. For $T>0$,
the decay rate becomes coupling-dependent except in the limit of high
couplings, where it reduces to a universal decay rate $\Gamma_{\varphi}=\pi T$.
These results were derived by a physically transparent semiclassical
approach that is exact in the high-energy limit. 

\emph{Acknowledgements}. \textendash{} We thank J. Chalker, Y. Gefen,
and O. Yevtushenko for fruitful discussions. Financial support by
DIP, NIM, the Emmy-Noether program and the SFB/TR 12 is gratefully
acknowledged.

\bibliographystyle{apsrev}
\bibliography{BibFM}

\end{document}